\documentclass[reprint,amsmath,amssymb,prl,aps,superscriptaddress]{revtex4-1} 
\usepackage{graphicx}
\usepackage{float}
\usepackage{physics}
\usepackage{amsthm}
\usepackage[utf8]{inputenc}
\usepackage[colorlinks]{hyperref}

\begin{document}
\title{Entangling Superconducting Qubits through an Analogue Wormhole}

%
\author{Carlos 
 Sab\'in}


\affiliation{
Instituto de F\'isica Fundamental, CSIC,
Serrano, 113-bis,
28006 Madrid, Spain; csl@iff.csic.es}





\begin{abstract}
We propose an experimental setup to test the effect of curved spacetime upon the extraction of entanglement from the quantum field vacuum to a pair of two-level systems. We consider two superconducting qubits coupled to a dc-SQUID array embedded into an open microwave transmission line, where an external bias can emulate a spacetime containing a traversable wormhole. We find that the amount of vacuum entanglement that can be extracted by the qubits depends on the wormhole parameters. At some distances qubits which would be in a separable state in flat spacetime would become entangled due to the analogue wormhole background.
\end{abstract}
\maketitle
\section{Introduction}
The vacuum of a quantum field is an entangled state~\cite{summerswerner, summerswernerII}. Vacuum fluctuations exhibit correlations between different space-time regions, even if they are spacelike separated. This fact underlies many important predictions of Quantum Field Theory in general backgrounds, such as the Dynamical Casimir Effect~\cite{moore} and Unruh--Hawking radiation~\cite{PhysRevD.7.2850,Hawking:1974, DaviesEffect, UnruhEffect, BellUnruh, BellUnruhReview}. From~a more applied viewpoint, it seems natural to ask if these correlations can be exploited as a resource for Quantum Information tasks.  This question can be addressed by two alternative approaches. On~the one hand, moving boundary conditions can turn vacuum fluctuations into real particles via the dynamical Casimir effect, which has recently been experimentally observed~\cite{casimirwilson}. These particles are produced in quantum-correlated pairs~\cite{dceentanglement, discord, ipsteering, coherence, casimirsimone}. On~the other hand, entanglement can in principle be swapped to qubits~\cite{reznik, reznikII, cqedsabin,past-future} after an interaction with the field. Despite several proposals, the~latter possibility has never been confirmed~experimentally.

In general, the~scenario for extracting vacuum entanglement can be described as follows. At~least two qubits are prepared in an uncorrelated state and interact for a finite time with a quantum field initially in the vacuum state. If~$t$ is the interaction time, $r$ is the distance between the qubits, and~$v$ is the propagation velocity of the field quanta, entanglement from the vacuum will be swapped to the qubits if their state is entangled after $t<r/v$. For~$t>r/v$, the~qubits might exchange real photons, which might act as an additional source of correlation. An~obvious experimental challenge is to achieve the desired interaction time, which requires control of the interaction on timescales that are typically out of reach. However, recent developments in the analysis of quantum information in relativistic scenarios show that the entanglement of relativistic quantum fields is sensitive to acceleration, gravity, and~the dynamics of spacetime~\cite{ivyalsingreview}. It seems natural to ask if we can exploit these properties to relax the experimental requirements necessary for extracting vacuum entanglement. Indeed, the~extraction of vacuum entanglement in curved spacetimes have been theoretically considered, for~instance in~\cite{educurved1,educurved2,educurved3}. However, for~experiments it is necessary to adopt an analogue gravity viewpoint~\cite{analoguereview2011, analogue1, analogue2, analogue3, analogue4} and search for experimental platforms where curved spacetimes can be~simulated.
 
Circuit QED 
~\cite{wallraff04,reviewnature} can be a natural framework to address the interaction of two-level systems with a quantum field. Superconducting qubits can be coupled to transmission lines, giving rise to an artificial one-dimensional matter-radiation interaction enjoying experimental accessibility and tunability of physical parameters. Exploiting these advantages, fundamental quantum-field problems typically considered as ideal can be accessible to experimental test. For~instance, the~ultrastrong coupling regime~\cite{bourassa09, forn-diaz10, peropadre10, niemczyk10, pollaquitas}  has already been leveraged in order to propose an experimental test of the vacuum entanglement extraction to a pair of spacelike separated artificial atoms~\cite{cqedsabin}. Moreover, effective spacetime metrics can be implemented as well by means of suitable modulations of the effective speed of light in the electromagnetic medium~\cite{UnruhColloquium}.

One important example of nontrivial curved background is the Ellis metric~\cite{ellis},  representing an spacetime which contains a traversable wormhole~\cite{morristhorne}. We do not have any experimental evidence of the presence of traversable wormholes in the universe, although~observational-based bounds have been determined~\cite{search}.  Indeed, the~existence of traversable wormholes would entail a challenge to the theoretical notion of causality~\cite{morristhorne,morristhorne2,hawking,deutsch}. This led Hawking to pose the ``chronology protection conjecture'' \cite{hawking}, which is formulated  within the semiclassical framework of quantum field theory in curved spacetime. According to this conjecture, quantum effects would prevent the creation of closed timelike curves in spacetimes such as Ellis, thus ruling out the possibility of time traveling to the past. The~conjecture could only be totally proved or disproved with a full theory of quantum gravity, which remains elusive. From~a strictly classical viewpoint, traversable wormholes would require exotic energy sources, namely, these sources would violate the weak energy condition~\cite{morristhorne2}. Moreover, quantum constraints can be inferred in the form of  ``quantum inequalities''~\cite{quantuminequalities}. However, as~unlikely as the existence of traversable wormholes might be, it is not completely forbidden on theoretical grounds. Furthermore, it has been suggested that phenomena typically attributed to black holes might as well be produced by exotic objects such as Ellis wormholes. If~wormholes existed, even~the origin of the detected gravitational waves could be questioned~\cite{gravastar,konoplya} together with the identity of the objects in the center of the galaxies~\cite{combi}. On~the other hand, the~existence of closed timelike curves would have a significant impact on classical and quantum computing~\cite{openctcs} and wormholes are at the core of the ``EPR-ER''
conjecture~\cite{maldacena}. For~all these reasons there is renewed interest in the theoretical characterization of these objects~\cite{geodesics, taylor,mapping} and in their detection by gravitational lensing~\cite{lensing, lensing2}, and~other classical methods~\cite{shadows}, or~by quantum techniques~\cite{qdwh,qdwhnonlinear}. Finally, classical~\cite{rousseax,rousseax2,magneticwormhole,nanophotonics} and quantum simulators of Ellis and other wormhole spacetimes have been discussed in quantum setups such as superconducting circuits~\cite{sabinwh}, trapped ions~\cite{mapping}, or~Bose--Einstein condensates~\cite{sabinmateos}.

In this paper, we consider a setup of two superconducting  qubits coupled to a dc-SQUID array embedded into an open transmission line, with~an external bias emulating the required traversable wormhole metric~\cite{sabinwh}.  We show that the vacuum entanglement that can be extracted by the two-level systems depends upon the wormhole parameters. The~features of this dependence are in turn sensitive to the distance between the qubits. We find a regime of distance where the presence of the wormhole entangles the artificial atoms, which would remain separable if the spacetime were~flat.

\section{Model and~Results}

A traversable 1-D section of a spacetime containing a massless traversable wormhole can be describd by the following line element~\cite{morristhorne2,sabinwh}:
\begin{equation}
ds^2=-c^2\,dt^2+\frac{1}{1-\frac{b(r)}{r}}\,dr^2, \label{eq:metric}
\end{equation}
where the shape function $b(r)$ encodes the wormhole features and is a function of the radius $r$ only. The~value $b_0$ of $r$ such that $b\, (r=b_0)=r=b_0$ determines the throat of the wormhole. The~proper radial position with respect to the throat is~\cite{morristhorne2} $l=\pm\int_{b_0}^r\,dr'(1-b(r')/r')^{-1/2}$, which defines two~different ``universes'' or branches in a single universe for $l >0$ ($r$ from $\infty$ to $b_0$) and $l<0$ ($r$ from $b_0$ to $\infty$). Therefore, as~$r\rightarrow\infty$ there are two asymptotically flat spacetime regions $l\rightarrow\pm\infty$, which are connected only by the throat of a wormhole in $l=0$ ($r=b_0$). 

An example of interest is~\cite{ellis,morristhorne, geodesics,taylor}:
\begin{equation}\label{eq:example}
b(r)=\frac{b_0^2}{r}, 
\end{equation}
which leads to $l^2(r)=r^2-b_0^2$. 

It is shown in~\cite{sabinwh} that the equations of motion of a quantum field in the spacetime given by Equation~(\ref{eq:metric}) is equivalent to the one in 
$ds^2=-c^2\,(1-\frac{b(r)}{r})\,dt^2+\,dr^2,$
where we can also define an r-dependent speed of propagation given by:
$c^2(r)=c^2\,(1-\frac{b(r)}{r})$. Furthermore, this effective speed of light can be mimicked for the electromagnetic flux field propagating in a dc-SQUID array embedded in an open transmission line~\cite{casimirsorin,array, array2,array3}, with~a particular  profile of the external field. In~particular, in~order to simulate the spacetime given by Equation~(\ref{eq:metric}) the profile must be:
\mbox{$\phi_{ext} (r)=\frac{\phi_0}{\pi}$ arccos $(1-\frac{b(r)}{r})$.} The~position $x$ along the transmission line can be related to the coordinate $r$ via $|x|=r-b_0,$ \mbox{$\quad x\in(-\infty,\infty)$}. Thus $x=0$ at the throat $r=b_0$ and has different signs at both sides. Notice that $l^2=|x|(|x|+2 b_0)$.
Thus, we can rewrite the flux profile as a function of $x$:
\mbox{$\phi_{ext} (x)=\frac{\phi_0}{\pi}$ arccos $(1-\frac{b_0^2}{(|x|+b_0)^2})$}. 
It is shown in~\cite{sabinwh} that in the particular wormhole spacetimes given by Equation~(\ref{eq:example}) it is possible to achieve a simulated wormhole throat radius in the sub-$\operatorname{mm}$ range. 

Now we assume that two superconducting qubits couple to the SQUID-embedded transmission line described above. The~qubit-line coupling strength can be abruptly switched on and off and can reach the ultrastrong coupling regime.  
In flat spacetime, the~extraction of entanglement from the quantum vacuum to a pair of superconducting qubits was analyzed in~\cite{cqedsabin}. The~Hamiltonian, $H = H_0 + H_I,$ splits into a free part $H_0$ for qubits and field
$H_0 = \frac{1}{2}\hbar\Omega(\sigma^z_A + \sigma^z_B) + \sum_k \omega(k)
  a^{\dagger}_ka_k$
and a standard interaction among them
$H_I \propto\sum_{\alpha=A,B} \sigma^x_A V(\chi_\alpha)$. $V$ is the quantum field, which is written in terms of the creation and annihilation operators
$ V(\chi)\propto\int dk \sqrt{N\omega_k}\left[e^{ik\chi}a_k +\mathrm{H.c.}\right]$. Here 
$\chi_A$ and $\chi_B$ would be the constant positions of the atoms in a coordinate system in which the spacetime metric is flat. In~the flat spacetime case of~\cite{cqedsabin} those are the standard laboratory coordinates $\{t,x\}$. In~the effective curved spacetime that we are considering here they are $\{t,l\}$.

We consider the initial state $|\psi(0)\rangle = |eg\rangle\otimes|0\rangle,$ with qubit $A$ in the excited state, while qubit $B$  is in its ground state and the field is in the vacuum state.  
The evolution in the interaction picture is given~by:
\begin{equation}
  \ket{\psi(t)} = {\cal T}[e^{-i \int_0^tdt' H_I(t')/\hbar}]\ket{eg}\otimes\ket{0},\label{c}
\end{equation}
where ${\cal T}$ is the time ordering operator. 
 We compute the corresponding two-qubit reduced density matrix $\rho_{12} $ after an interaction time $t$ in second-order perurbation theory beyond the Rotating Wave Approximation and tracing over the field~\cite{cqedsabin}. The~amount of entanglement of the X-state can be computed by several means. We choose one of the most standard measures for two-qubit systems, namely the concurrence, which is: $ \mathcal{C}(\rho_{12})=2 \Big[|X|-\Big({\sum_{k}|A_{1,k}|^2 \sum_{k}|B_{1,k}|^2}\Big)^{1/2} \Big]$, where $X$ stands for the amplitude of photon exchange and $\sum_{k}|A_{1,k}|^2$,  $\sum_{k}|B_{1,k}|^2$ for the probability of emission of a photon by qubits $1$ and $2$, respectively. Using the techniques in~\cite{cqedsabin} we compute these magnitudes as a function of three dimensionless parameters, $\xi$, $K_1$, and~$K_2$.  The~first one, $\xi=c\,t/\rho$, ($\rho$ being the constant distance between the qubits $\rho=|\chi_a - \chi_b|$) allows us to discriminate between two different spacetime regions, namely, the~qubits are effectively spacelike separated if $\xi<1$ and timelike separated otherwise. The~remaining parameters are the dimensionless coupling strengths for qubits 1 and 2:  $K_m= \left(g_m/\Omega_m\right)^2.$ We will restrict our analysis to $K_m\Omega_m\,t \ll 1 $ where our perturbative approach remains valid.  We assume that $g_1=g_2=g$ and $\Omega_1=\Omega_2=\Omega$, and~thus $K_1=K_2=K$.

The results of~\cite{cqedsabin} are directly valid to the coordinate system $\{t,l\}$, for~which the metric is flat. It~is therefore necessary to transform the parameter $\xi$ to the laboratory coordinates. For~simplicity, we~assume that the qubit position are symmetric with respect to the throat:
\begin{equation}\label{eq:qubitpos}
x_B=|x_A|=-x_A.
\end{equation}

Then, by~using the relation between $x$ and $l$:
\begin{equation}\label{eq:distanceflat}
l_B-l_A= 2 x_B \sqrt{1+2\xi_b},
\end{equation}
where we introduce a dimensionless parameter:
\begin{equation}\label{eq:chib}
\xi_b=\frac{b_0}{x_B},
\end{equation}
relating the throat size with the qubit distance in the laboratory coordinates $\rho_x= 2x_B$. Thus,~from~Equation~(\ref{eq:distanceflat}) we get the relation between the qubit distance in free-falling and laboratory~coordinates:
\begin{equation}
\label{eq:relationdistance}
\rho_l=\rho_x \sqrt{1+2\xi_b}.
\end{equation}

Moreover, while in the flat coordinate system the time that takes the light to travel between qubits is merely $(l_B-l_A)/c$, in~the laboratory this time will be given by:
\begin{eqnarray}\label{eq:labqubittime}
t_{AB}&=&\int_{x_A}^{x_B}\frac{dx}{c(x)}=\frac{l_B-l_A}{c}-2\frac{b_0}{c}\operatorname{arcsinh}(\frac{1}{\sqrt{2\xi_B}})+\nonumber\\ & &\frac{b_0}{c}\operatorname{log}\left(1+\frac{1}{\xi_B}(1+\sqrt{1+2\xi_B})\right).
\end{eqnarray}

Thus, the~new parameter:
\begin{equation}\label{eq:xilab}
\xi_x=\frac{t}{t_{AB}}
\end{equation}
will define the light cone in the laboratory system. If~$\xi_x<1$, light cannot travel between the qubits. By~means of Equation~(\ref{eq:labqubittime}), we can relate $\xi_x$ with the light cone parameter in free-falling coordinates:
\begin{equation}\label{eq:xil}
\xi_l=\frac{c t}{l_B-l_A}.
\end{equation}

We get:
\begin{eqnarray}\label{eq:chilchix}
&&\xi_l=\\
&&\frac{1}{\frac{1}{\xi_x}+\frac{\xi_B}{\xi_F}\operatorname{arcsinh}(\frac{1}{\sqrt{2\xi_B}})-\frac{\xi_B}{2 \xi_F}\operatorname{log}\left(1+\frac{1}{\xi_B}(1+\sqrt{1+2\xi_B})\right)},\nonumber
\end{eqnarray}
where we introduce $\xi_F$, which would be the light cone parameter in flat spacetime:
\begin{equation}\label{eq:chif}
\xi_F=\frac{c t}{2 x_B}.
\end{equation}

\begin{figure}[H]\
\includegraphics[width=.50\textwidth]{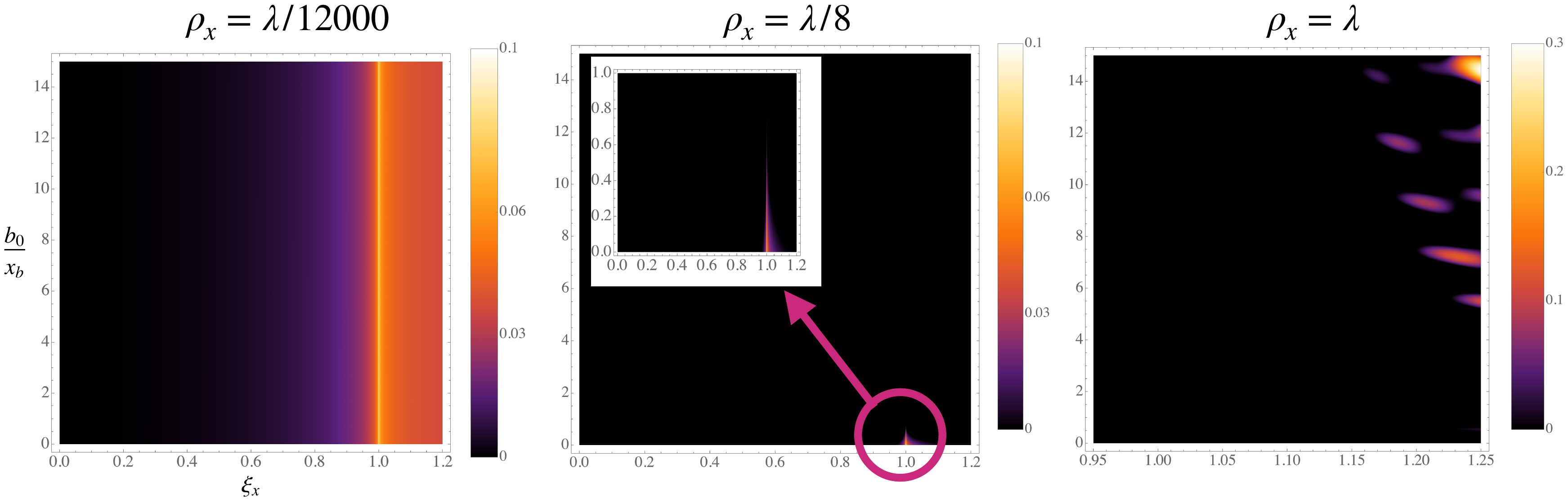} 
\caption{Concurrence vs. the light-cone parameter $\varepsilon_x$ and the wormhole parameter $\varepsilon_b$ for three~different qubit distances and $K=7.5\cdot10^{-3}$. For~the shortest distance (\textbf{left}), the~effect of the presence of the analogue wormhole is mostly irrelevant. For~medium distances (\textbf{middle}), entanglement appears in the flat-spacetime case and quickly vanishes as $b_0$ grows (see the inset).~For~larger distances (\textbf{right}), we~find the opposite scenario: There is entanglement only in the presence of the effective~wormhole.} \label{fig1}
\end{figure}

Note that in the absence of a wormhole ($b_0=0$), from~Equation~(\ref{eq:chilchix}) we get $\xi_l=\xi_x=\xi_F$, as~expected. 
Generally speaking, it is expected that the extraction of entanglement decay with the distance mimicking the decay of vacuum correlations of the field.
By inserting Equations~(\ref{eq:relationdistance}) and (\ref{eq:chilchix}) in the flat-spacetime results, we show in Figure~\ref{fig1} the dependence of the entanglement dynamics on the wormhole parameter $b_0$. As~expected, it is highly dependent on the qubit distance, since vacuum correlations decay with distance. Indeed, we find three separate regimes. For~ultrashort qubit distances $\rho_x\ll \lambda$---where $\lambda=2\pi c/\Omega$ would be the wavelength of the qubit transition in flat spacetime---there is entanglement between qubits both inside and outside the light cone which seems to be independent of the existence and size of a wormhole throat. As~$\rho_x$ is increased the dynamics of entanglement becomes sensitive to the wormhole. First, the~effect is detrimental: At distances at which there is entanglement generation around the light cone in the absence of a wormhole---in agreement with the flat-spacetime results~\cite{cqedsabin}---entanglement vanishes quickly as $\varepsilon_b$ increases. However, for~larger distances $\rho_x\simeq\lambda$ some entanglement is generated for timelike separated qubits only if $\varepsilon_b\neq0$. Therefore, qubits which would be in a separable state in flat spacetime would get entangled due to the existence of the curved background. Since $\varepsilon_x>1$, we cannot say that this is a pure transference of vacuum entanglement, since~there could have been photon exchange between the qubits. However, it is still interesting that the effect is a consequence of the presence of an effective curved spacetime. The~topological link between the qubits provided by the wormhole enhances photon exchange, making it strong enough to generate quantum~correlations.
This entanglement generation between qubits due to an analogue curved spacetime is the main result of this work. We believe that it is within reach of circuit QED technology. Superconducting transmon qubits has been coupled to transmission lines formed by thousands of SQUIDS~\cite{qubitandarray}. In~\cite{sabinwh} we showed that it is possible to simulate wormholes with $b_0$ in the sub-mm range by means of an inhomogenuous external magnetic field bias. On~the other hand, we need $b_0$ to be slightly larger than the wavelength $\lambda$, since $b_0/\lambda\simeq b_0/\rho_x=b_0/(2x_b)=\varepsilon_b/2$ and interesting effects show up for $\varepsilon_b\geq5$. This means that $\lambda$ in the sub-mm range is required. For~qubits of $\Omega=2\pi\times10$ GHz this means propagation speed velocities $c$ for the field  of around $10^6 \operatorname{m/s}$. These speeds has been experimentally reported  in SQUID arrays~\cite{apl}.  Another experimental challenge would be the preparation of a pure initial quantum vacuum for the field. However, for~realistic temperatures of 30 $\operatorname{mK}$ 
 \cite{threephotons} the number of thermal photons at 5 $\operatorname{mK}$ is as low as $3\cdot 10^{-4}$ so the effect of thermal noise should be negligible. In~order to test the effects of effective curved spacetime on the extraction of vacuum entanglement, the~experiment should be repeated with and without external bias (wormhole) and the comparison of the results should follow our predictions. A~fully detailed experimental description would lie beyond the scope of the current theoretical work. 
\section{Conclusions}

In summary, we proposed a setup of two superconducting qubits coupled to a SQUID array transmission line to test the extraction of entanglement from the vacuum of a quantum field in curved spacetime, in~particular, the~Ellis wormhole metric, simulated by means of a suitable external bias of the SQUID array. It was found that different regimes in which, according to the distance between the qubits, the~presence of the Ellis wormhole could be irrelevant, detrimental to entanglement generation, or~beneficial. We think that the latter was the most interesting scenario. In~particular, it was found that when the distance between the qubits was similar to the wavelength of the qubit transition, there~was no entanglement extraction unless the wormhole throat was significantly larger than a certain value ($\varepsilon_b\geq5$). This means that a pair of qubits which would be separable in flat spacetime would become entangled due to the effective curved background. We have shown that this would be in principle observable with current circuit QED technology. Therefore, we found that the analysis of analogue quantum simulators of quantum field theory in curved spacetime was not only interesting from a theoretical viewpoint and has possible experimental applications as a valuable resource for quantum~technologies.




\vspace{6pt} 




\section*{Acknowledgements} C.S. has received financial support through the Postdoctoral Junior Leader Fellowship Programme from la Caixa Banking Foundation (LCF/BQ/LR18/11640005).









\end{document}